\documentclass[
 aps,prl,showpacs,superscriptaddress,numerical,amsmath,amssymb,floatfix,
reprint,
]{revtex4-1}
\usepackage{xcolor}
\usepackage[
	pdffitwindow=true,
	colorlinks=true,
	frenchlinks=false,
        linkcolor=blue,
	anchorcolor=blue,
        citecolor=blue,
        filecolor=blue,
        urlcolor=blue,
        bookmarks=true,
        bookmarksopen=true,
	bookmarksnumbered=true,
        bookmarksopenlevel=1,
        plainpages=false,
	pdfpagelayout=TwoPageLeft,
        pdfpagelabels=true,
	breaklinks
]{hyperref}
\usepackage[per-mode=symbol,separate-uncertainty]{siunitx}
\usepackage{graphicx}
\usepackage{dcolumn}
\usepackage{bm}
\usepackage[english]{babel}
\usepackage{color}
\usepackage{ulem}

\begin{document}
\title[]{Spin transport in a magnetic insulator with zero effective damping}

\author{T.~Wimmer}
\email[]{tobias.wimmer@wmi.badw.de}
\affiliation{Walther-Mei{\ss}ner-Institut, Bayerische Akademie der Wissenschaften, 85748 Garching, Germany}
\affiliation{Physik-Department, Technische Universit\"{a}t M\"{u}nchen, 85748 Garching, Germany}
\author{M.~Althammer}
\email[]{matthias.althammer@wmi.badw.de}
\affiliation{Walther-Mei{\ss}ner-Institut, Bayerische Akademie der Wissenschaften, 85748 Garching, Germany}
\affiliation{Physik-Department, Technische Universit\"{a}t M\"{u}nchen, 85748 Garching, Germany}
\author{L.~Liensberger}
\affiliation{Walther-Mei{\ss}ner-Institut, Bayerische Akademie der Wissenschaften, 85748 Garching, Germany}
\affiliation{Physik-Department, Technische Universit\"{a}t M\"{u}nchen, 85748 Garching, Germany}
\author{N.~Vlietstra}
\affiliation{Walther-Mei{\ss}ner-Institut, Bayerische Akademie der Wissenschaften, 85748 Garching, Germany}
\author{S.~Gepr{\"a}gs}
\affiliation{Walther-Mei{\ss}ner-Institut, Bayerische Akademie der Wissenschaften, 85748 Garching, Germany}
\author{M.~Weiler}
\affiliation{Walther-Mei{\ss}ner-Institut, Bayerische Akademie der Wissenschaften, 85748 Garching, Germany}
\affiliation{Physik-Department, Technische Universit\"{a}t M\"{u}nchen, 85748 Garching, Germany}
\author{R.~Gross}
\affiliation{Walther-Mei{\ss}ner-Institut, Bayerische Akademie der Wissenschaften, 85748 Garching, Germany}
\affiliation{Physik-Department, Technische Universit\"{a}t M\"{u}nchen, 85748 Garching, Germany}
\affiliation{Nanosystems Initiative Munich (NIM), Schellingstra{\ss}e 4, 80799 M\"{u}nchen, Germany}
\affiliation{Munich Center for Quantum Science and Technology (MCQST), Schellingstr. 4, D-80799 M\"{u}nchen, Germany}
\author{H.~Huebl}
\email[]{hans.huebl@wmi.badw.de}
\affiliation{Walther-Mei{\ss}ner-Institut, Bayerische Akademie der Wissenschaften, 85748 Garching, Germany}
\affiliation{Physik-Department, Technische Universit\"{a}t M\"{u}nchen, 85748 Garching, Germany}
\affiliation{Nanosystems Initiative Munich (NIM), Schellingstra{\ss}e 4, 80799 M\"{u}nchen, Germany}
\affiliation{Munich Center for Quantum Science and Technology (MCQST), Schellingstr. 4, D-80799 M\"{u}nchen, Germany}

\date{\today}


\begin{abstract}
Applications based on spin currents strongly profit from the control and reduction of their effective damping and their transport properties. We here experimentally observe magnon mediated transport of spin (angular) momentum through a $\SI{13.4}{\nano\metre}$ thin yttrium iron garnet film with full control of the magnetic damping via spin-orbit torque. Above a critical spin-orbit torque, the fully compensated damping manifests itself as an increase of magnon conductivity by almost two orders of magnitude. We compare our results to theoretical expectations based on recently predicted current induced magnon condensates and discuss other possible origins of the observed critical behaviour.
\end{abstract}
\maketitle

\textit{Introduction} - There is broad interest in using the spin degree of freedom for information transport. This makes the efficient manipulation of spin currents an important but also challenging task~\cite{Ahopelto2019,Jansen2012,Hoffmann2007,Coll2019}. Magnons, the quantized excitations of the spin system in a magnetically ordered material, are one of the most promising candidates for the transport of spin information. However, in contrast to the number of charge carriers in an electronic conductor, the magnon number in a spin conductor is not conserved. Inevitably, magnon mediated spin currents only prevail on a characteristic length scale, which is mainly determined by the magnetic Gilbert damping of the material. Therefore, efficient ways of reducing and tuning the magnetic damping represent an important step for spin transport devices. 

One possible way to manipulate spin currents is to employ spin orbit torques (SOTs) in heavy metal (HM)/ferromagnetic insulator (FMI) bilayers~\cite{Klein2014,Collet2016,Evelt2016,Evelt2018}. Driving a charge current through the HM in contact with the FMI, an antidamping-like spin torque can be exerted on the magnetization of the FMI. Above a critical current, the magnetic damping is completely compensated via the SOT. For nano-structured devices, this damping compensation manifests itself in the emergence of auto-oscillations of the magnetization~\cite{Tsoi1998,demidov_magnetic_2012,Klein2014,Collet2016}. Previous experiments~\cite{Evelt2016} demonstrated a 10-fold increase of the propagation length of coherent spin-waves in a HM/FMI waveguide upon application of a large charge current to the HM. Cornelissen et al~\cite{LudoTransistor} reported that also the diffusive transport of incoherently generated magnons can be controlled by charge currents in HM/FMI nanostructures.



\begin{figure}[]%
	\includegraphics[]{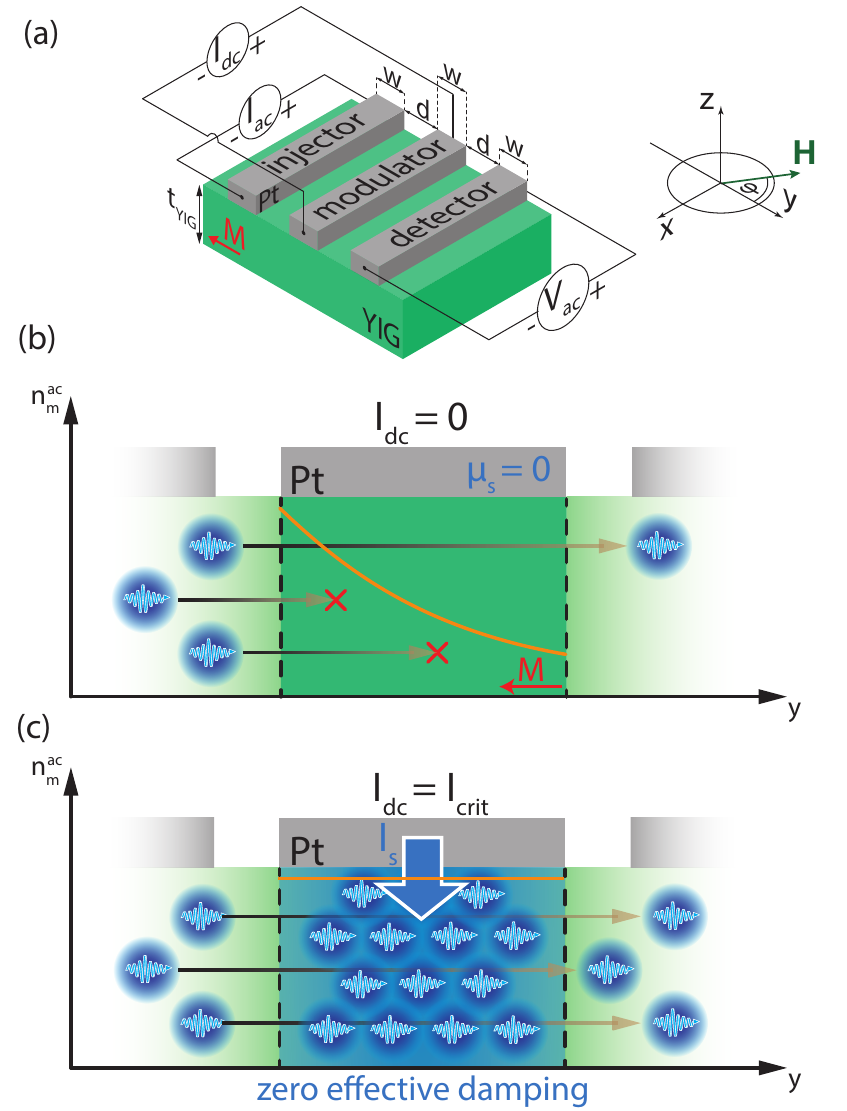}%
	\caption{(a) Schematic depiction of the device, electrical connection scheme and the coordinate system with the in-plane rotation angle $\varphi$ of the applied magnetic field $\mu_0 H$. 
		(b), (c) Illustrations of the magnon transport from injector to detector. We here only consider magnon transport directly below the modulator. (b) For $I_\mathrm{dc}=0$, magnons (blue wiggly arrows) generated by the injector diffuse from left to right. Magnon decay events, indicated by red crosses, result in a finite lifetime and a corresponding characteristic spin diffusion length depicted as a exponential decay of the magnon density $n_\mathrm{m}^\mathrm{ac}$ (orange solid line). The modulator only statically affects the transport properties via magnon absorption. (c) For $I_\mathrm{dc} = I_\mathrm{crit}$, the modulator current is large enough to compensate the magnetic damping of the YIG, resulting in effectively vanishing magnon decay beneath the modulator. The damping compensation is illustrated by a large magnon accumulation beneath the modulator. 
	}%
	\label{fig:scheme}%
\end{figure}

In this Letter, we demonstrate the full compensation of the magnetic damping in a nanometer-thick yttrium iron garnet (YIG) film via SOT caused by a charge current in an adjacent HM layer. Above a threshold current density in the HM, we observe a highly non-linear increase of magnon conductivity by almost two orders of magnitude, indicating vanishing magnon decay. Our experimental observations can be rationalized by a SOT induced damping compensation of the magnetization dynamics. In this context, we will discuss two possible scenarios leading to the damping compensation: (i) a strong overpopulation of modes by incoherent magnons and (ii) the formation of a coherent auto-oscillation state~\cite{Collet2016} equivalent to a swasing state~\cite{dcmagnonBEC}. In addition to (ii), Bender et al. predict the formation of a magnon Bose-Einstein condensate (BEC) for SOT levels below the swasing phase. The onset and smooth transition of the observed change in the magnon conductance might be indicative for this BEC phase~\cite{Demokritov2006,dcmagnonBEC}.



\textit{Observation} - The principle of our magnon conductance measurement is inspired by recent DC magnetotransport experiments that infer magnon transport properties in YIG~\cite{CornelissenMMR,SchlitzMMR,KathrinLogik,KleinMMR,CornelissenTheory,Althammer2018,Shan2017,
Li2016,Klaui2018}. As shown in Fig.~\ref{fig:scheme} (a), magnons are injected from a Pt strip (injector) into a $\SI{13.4}{\nano\metre}$ thick YIG film by the spin Hall effect (SHE)~\cite{HirschSHE,Dyakonov} using a low-frequency ($\SI{13}{\hertz}$) charge current $I_\mathrm{ac} = \SI{50}{\micro\ampere}$ in the Pt strip. The diffusive transport of these magnons is quantified by electrically measuring the magnon density below a second Pt strip (detector) as the first harmonic voltage signal $V_\mathrm{ac}$ via lock-in detection, exploiting the inverse SHE (we plot one quadrature containing the entire lock-in signal). Cornelissen \textit{et al.}~\cite{LudoTransistor} demonstrated that the magnon transport in such an arrangement can be controlled by a DC charge current $I_\mathrm{dc}$ applied to a third (modulator) strip placed in between injector and detector (c.f.~Fig.~\ref{fig:scheme} (a)). 
The modulator current causes a finite spin chemical potential $\mu_\mathrm{s}$ at the Pt/YIG interface, leading to an enhanced magnon density in YIG. Since the magnon chemical potential $\mu_\mathrm{m}$ is expected to grow with $\mu_\mathrm{s}$, we can tune $\mu_\mathrm{m}$ by varying $I_\mathrm{dc}$. 

In contrast to Ref.~\cite{LudoTransistor}, we here focus on the non-linear regime of this magnon transport. Our physical picture of the magnon transport is condensed in Fig.~\ref{fig:scheme} (b) and (c). For the sake of simplicity, we only consider magnon transport beneath the modulator and therefore disregard the magnon decay on either side of the modulator. When $I_\mathrm{dc} = 0$ (panel (b)), the magnon density $n_\mathrm{m}^\mathrm{ac}$ from the injector decays exponentially (orange solid line). For $I_\mathrm{dc} = I_\mathrm{crit}$ (panel (c)), the threshold current for the damping compensation is reached, the magnon lifetime diverges and spin transport with an effectively vanishing magnon decay ensues. This corresponds to a zero effective damping state and is illustrated by the large magnon accumulation beneath the modulator. 
\begin{figure}[]%
	\includegraphics[]{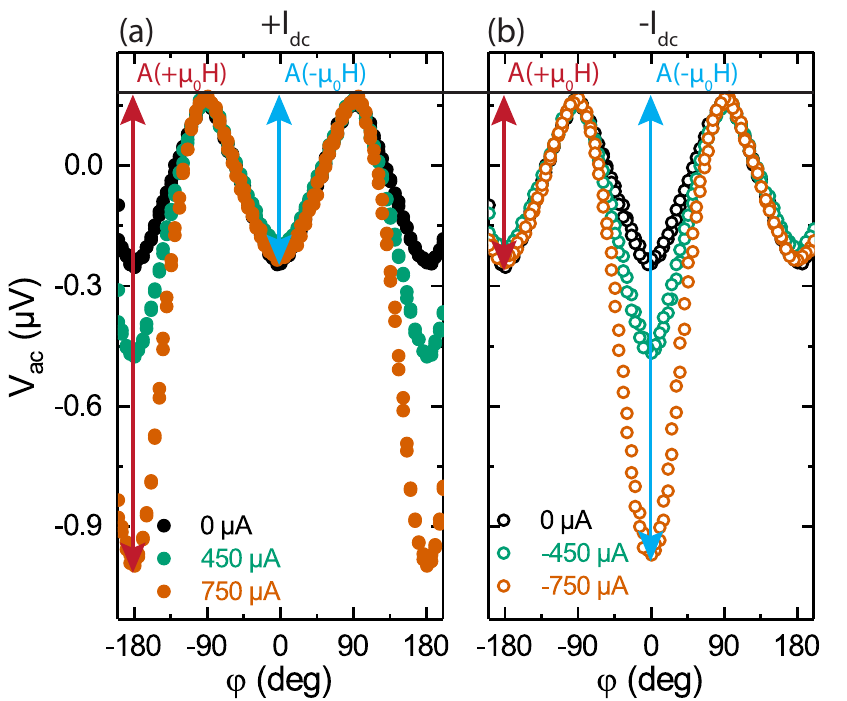}%
	\caption{Detector signal $V_\mathrm{ac}$ plotted versus the rotation angle $\varphi$ of the in-plane field at $\mu_0 H = \SI{50}{\milli\tesla}$ for (a) positive and (b) negative DC bias currents $I_\mathrm{dc}$ in the modulator. (a) The magnon transport signal for $I_\mathrm{dc} > 0$ is significantly increased at $\varphi = \pm\SI{180}{\degree}$ and mostly unaffected at $\varphi = \SI{0}{\degree}$. (b) For $I_\mathrm{dc} < 0$, we observe a $\SI{180}{\degree}$ shifted behavior, where the signal increase is evident at $\varphi = \SI{0}{\degree}$, while unchanged for $\varphi = \pm\SI{180}{\degree}$. 
	}%
	\label{fig:rawdata}%
\end{figure}

To investigate the magnon propagation in the thin YIG layer for different modulator currents $I_\mathrm{dc}$
, we measure $V_\mathrm{ac}$ as a function of the magnetic field orientation $\varphi$ (c.f.~Fig.~\ref{fig:scheme} (a)) with a fixed magnetic field strength of $\mu_0 H = \SI{50}{\milli\tesla}$ at $T=\SI{280}{\kelvin}$. The result is shown in Fig.~\ref{fig:rawdata}, 
where the black data points show the characteristic ($\cos^2{\varphi}$) modulation expected for magnon transport between injector and detector for $I_\mathrm{dc}=0$. This results from the variation of the magnon injection with $\varphi$, with maxima expected for $\mathbf{H}$ perpendicular to $I_\mathrm{ac}$ ($\varphi=\SI{-180}{\degree}, \SI{0}{\degree}, \SI{180}{\degree}$)~\cite{CornelissenMMR,SchlitzMMR}. Note that we observe a finite offset signal even at $\varphi=\pm\SI{90}{\degree}$. Since this offset signal is found to be non-reproducible in different measurement setups, we attribute this to a spurious experimental artifact. The rather triangular shape of the angle dependent measurement for $I_\mathrm{dc} = 0$ is due to the cubic magnetocrystalline anisotropy of the YIG film (see Ref.~\footnotemark[1]), which results in non-collinear orientations of the magnetization $\mathbf{M}$ and the external field $\mathbf{H}$. 
Most importantly, however, a significant enhancement of the magnon transport signal is observed at $\varphi = \pm\SI{180}{\degree}$ in Fig.~\ref{fig:rawdata} (a) for $I_\mathrm{dc}>0$. 
This can be understood by a magnon accumulation underneath the modulator, which increases the magnon conductivity and results in a larger $V_\mathrm{ac}$. In the same way, a decrease of $V_\mathrm{ac}$ is expected for $\varphi=\SI{0}{\degree}$ due to the magnon depletion obtained in this configuration. This, however, is counterbalanced by thermally injected magnons present due to Joule heating of the modulator strip. Figure~\ref{fig:rawdata} (b) shows the measurement for the inverted DC current direction ($I_\mathrm{dc} < 0$). Here, we observe the expected $\SI{180}{\degree}$ shifted case: an enhancement for $\varphi = \SI{0}{\degree}$ and no significant change for $\pm\SI{180}{\degree}$. This behaviour is consistent with an accumulation of magnons for the given current and magnetic field direction.

\begin{figure}[!h]%
	\includegraphics[]{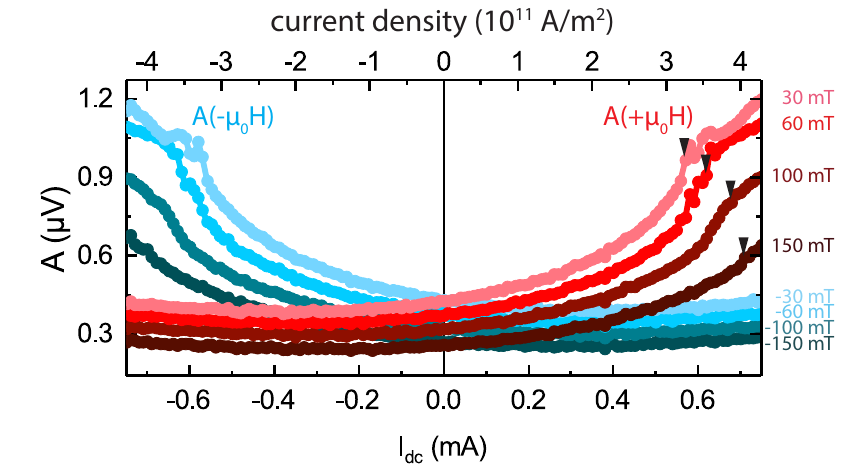}%
	\caption{Extracted amplitudes $A(+\mu_0 H)$ and $A(-\mu_0 H)$ (as indicated in Fig.~\ref{fig:rawdata}) of the magnon transport signal for different external magnetic fields plotted versus the DC current $I_\mathrm{dc}$ in the modulator. The transition into the damping compensation state for positive $I_\mathrm{dc}$ is indicated by black triangles (maximum slope of the curves). The transition shifts to larger DC currents with increasing external magnetic fields. 
	}%
	\label{fig:currentsweeps_v2}%
\end{figure}

For a quantitative analysis of the data presented in Fig.~\ref{fig:rawdata}, we extract the signal amplitudes $A(+\mu_0 H)$ and $A(-\mu_0 H)$ as a function of $I_\mathrm{dc}$ for various magnetic field amplitudes $H$ (see Fig.~\ref{fig:currentsweeps_v2}). 
In the low bias regime ($|I_\mathrm{dc}| < \SI{0.4}{\milli\ampere}$ ), the $A(I_\mathrm{dc})$ curves can be modeled by a superposition of a linear and quadratic dependence as already reported by Cornelissen et al.~\cite{LudoTransistor}. However, we observe a two orders of magnitude improved control of the magnon conductivity compared to Ref.~\cite{LudoTransistor}. This is in agreement with the predicted magnetic layer thickness dependence of the modulation efficiency~\cite{LudoTransistor}. A quantitative comparison to the model of Ref.~\cite{LudoTransistor} is shown in the Supplemental Material (SI)~\footnote{\label{fn:supp}See Supplemental Material at [url], which includes Refs.~\cite{Brataas2002,Zhang2015,Tserkovnyak2002,Zwierzycki2005,Barati2014,Kaplan1965,XiaoSSE,SchreierSSE,Thiery2018,CornelissenField,Guo2018,Stiles2006,Stancil2009,AltiSMR,Ando2009,Costache2006,CzeschkaSP,Liu2017,VanWeesThickness,Flebus2016}, for details on sample preparation, the determination of the material parameters, the impact of the spin Seebeck effect on our measurements, thermometry, injector-detector separation dependent measurements, the evaluation of the YIG spin resistance, the impact of magnetocrystalline anisotropy and the damping compensation on angular dependent signals, a thorough comparison of a DC pumped magnon BEC and spin torque oscillators and additonal experiments using microwave driven excitation of the magnetization.}. 
In addition, and most importantly, we see a pronounced deviation from the linear transport modulation~\cite{LudoTransistor} for large $I_\mathrm{dc}$. This manifests itself by a shoulder in the $A(I_\mathrm{dc})$ curves for $I_\mathrm{dc} > \SI{0.5}{\milli\ampere}$ (marked by black triangles in Fig.~\ref{fig:currentsweeps_v2} for positive $I_\mathrm{dc}$).

\begin{figure}[]%
	\includegraphics[]{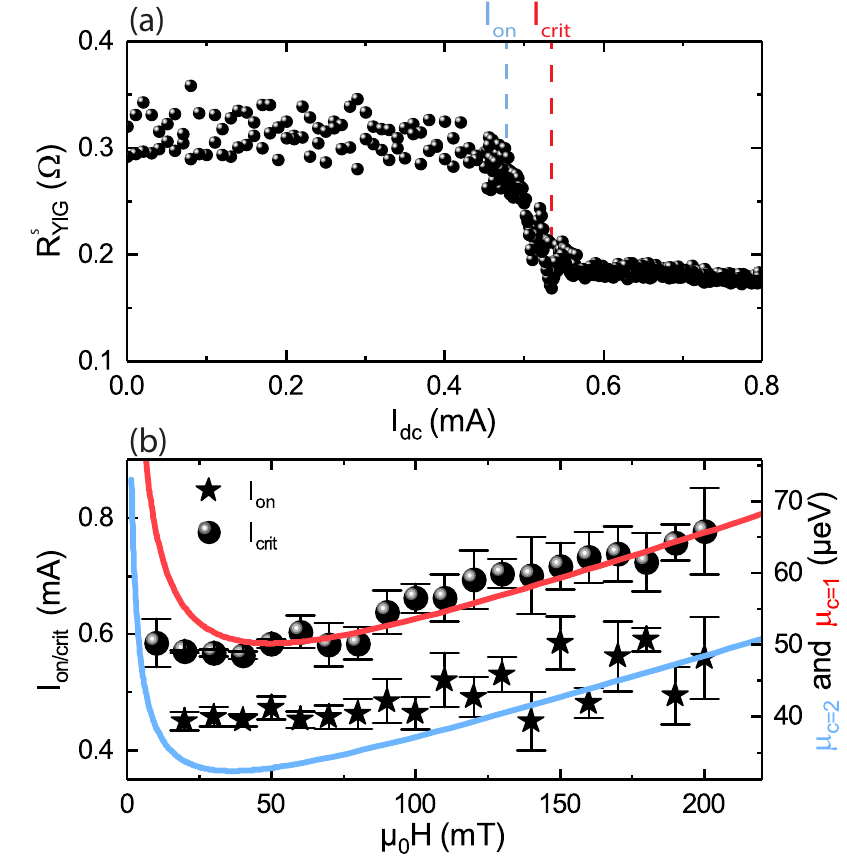}%
	\caption{(a) Magnon resistance $R_\mathrm{YIG}^\mathrm{s}$ of the YIG channel between injector and detector for a magnetic field of $\mu_0 H = \SI{50}{\milli\tesla}$. $R_\mathrm{YIG}^\mathrm{s}$ is corrected for effects associated with (linear) SHE and (quadratic) thermal magnon injection effects. A very steep decrease of $R_\mathrm{YIG}^\mathrm{s}$ for $I_\mathrm{on}<I_\mathrm{dc}<I_\mathrm{crit}$ is evident. The reduction of $R_\mathrm{YIG}^\mathrm{s}$ by $\SI{0.13}{\ohm}$ 
	is compatible with a vanishing magnon resistivity underneath the modulator strip. (b) Critical currents $I_\mathrm{on/crit}$ versus applied field $\mu_0 H$. 
	The right $y$-axis shows the critical chemical potentials $\mu_\mathrm{c=2/c=1}$ from Eq.~\eqref{eq:threshold} (solid red and blue lines). 
	}
	\label{fig:threshold}%
\end{figure}

We now focus on the magnon transport properties, which we express by an effective magnon resistance $R_\mathrm{YIG}^\mathrm{s}$. To this end, we evaluate $R_\mathrm{YIG}^\mathrm{s}$ measured between injector and detector as a function of the modulator current. 
The magnon resistance in YIG can be directly deduced from the magnon transport amplitudes $A$ plotted in Fig.~\ref{fig:currentsweeps_v2}~(see SI \footnotemark[1]). However, $A$ contains contributions from thermal (quadratic in $I_\mathrm{dc}$) as well as SHE induced magnon injection effects (linear in $I_\mathrm{dc}$). We correct for both of those contributions, leading to the $R_\mathrm{YIG}^{\mathrm{s}}(I_\mathrm{dc})$ dependence shown in Fig.~\ref{fig:threshold} (a) (for details see Ref.~\footnotemark[1]).
Thus, $R_\mathrm{YIG}^\mathrm{s}(I_\mathrm{dc})$ enables us to determine the impact on magnon transport stemming solely from non-linear and non-quadratic modulations of the magnon transport, i.e.~from the damping compensation regime. 
For $I_\mathrm{dc} < \SI{0.4}{\milli\ampere}$, we observe a constant $R_\mathrm{YIG}^\mathrm{s}$. We define a characteristic onset current $I_\mathrm{on}$, at which the magnon resistance $R_\mathrm{YIG}^\mathrm{s}$ starts to drop rapidly by $\SI{0.13}{\ohm}$ and saturates at a finite value above the second characteristic current $I_\mathrm{crit}$. 
Here, we define $I_\mathrm{on}$ as the current at which $R_\mathrm{YIG}^\mathrm{s}$ drops by $\SI{10}{\percent}$ compared to the constant resistance observed for small $I_\mathrm{dc}$. $I_\mathrm{crit}$ is taken at the current level where $R_\mathrm{YIG}^\mathrm{s}$ reaches its minimum value. 
The magnon resistance data also allows us to roughly estimate the resistance within the damping compensated region beneath the modulator. Assuming a serial resistor network model~\cite{CornelissenTheory} (see also the SI~\footnotemark[1]), and zero magnon resistance underneath the modulator strip when the damping is compensated, we expect $R_\mathrm{YIG}^\mathrm{s}=\SI{0.19}{\ohm}$ 
for $I_\mathrm{dc}>I_\mathrm{crit}$. This in good agreement with our data shown in Fig.~\ref{fig:threshold} (a). 
We can further roughly estimate the magnon resistivity $\rho_\mathrm{YIG}^\mathrm{s}$ for $I_\mathrm{dc}>I_\mathrm{crit}$ and obtain $
\SI{8.16}{\nano\ohm\metre}$, which is almost two orders of magnitude smaller than the magnon resistivity for $I_\mathrm{dc}<I_\mathrm{on}$ ($\SI{0.54}{\micro\ohm\metre}$)~\footnotemark[1]. Thus, the observed magnon resistance shows similarities to the sudden electrical resistance drop of a superconductor at the superconducting phase transition. 

\textit{Interpretation} - Our magnon transport measurements show that the magnon conductance is strongly enhanced for large $I_\mathrm{dc}$, i.e.~when the damping is compensated underneath the modulator strip. 
The strong enhancement suggests a vanishing magnon resistivity, which could be interpreted as spin superfluidity~\cite{Hillebrands2016,Brataas2015,Brataas2017,Takei2014}. As the damping compensation may also lead to coherent magnetization dynamics, this warrants the question how a coherent magnetization state created by ferromagnetic resonance (FMR) affects the transport properties. In stark contrast to the reduction of the magnon resistance due to the damping compensation, we find an increase of the magnon resistance when coherently driving the YIG magnetization by a microwave magnetic field~\footnotemark[1]~\cite{Liu2018}. This demonstrates that the effective compensation of magnetic damping is responsible for the formation of the ultra-low magnon resistance state - and not the coherence of the magnetization precession. In particular, we want to emphasize that damping compensation not necessarily results in a coherent precession of the magnetization, but also a broad frequency spectrum of excited modes is a possible scenario. 
Thus, taking the damping compensation as the bottom line of our experimental observations, we provide two possible scenarios explaining our findings: (i) a strong overpopulation of magnons in a broad frequency spectrum leads to the compensation of the magnon damping, but no coherent magnetization precession is achieved. (ii) Similar to a spin-torque-oscillator~\cite{Collet2016}, the compensation of the magnetic damping leads to a coherent auto-oscillation state of the magnetization, equivalent to a swasing phase as discussed below. Here, the terminology of swasing is adopted from Ref.~\cite{dcmagnonBEC} describing the spin wave analogon of lasing~\cite{Berger1996}.

In the following, we will compare our data to Ref.~\cite{dcmagnonBEC}, which theoretically predicts magnon condensation and swasing under DC pumping~\cite{dcmagnonBEC}. We want to emphasize that this swasing instability is identical to the threshold for auto-oscillations in spin Hall oscillators, as observed in Refs.~\cite{Tsoi1998,demidov_magnetic_2012,Klein2014,Collet2016} (see~\footnotemark[1] for a thorough derivation of this equivalence). Note, that this threshold condition is independent of the scenario and thus also holds for the incoherent case (i), since damping compensation is given by the equality of the magnon relaxation and pumping rate and hence assumes no coherence of the excited modes.  
This corresponds to the case $c=1$ in the subsequent discussion (see Eq.~\eqref{eq:threshold}). In addition, the model by Bender et al.~\cite{dcmagnonBEC} also discusses the formation of a magnon BEC ($c=2$ in Eq.~\eqref{eq:threshold}), which is defined by a finite population of magnons in the ground state. 
We rewrite their model to conform to our in-plane magnetized case~\footnotemark[1] and find the spin chemical potentials $\mu_\mathrm{c=2}$, corresponding to the formation of a magnon BEC, as well as $\mu_\mathrm{c=1}$, i.e.~the so-called swasing instability in the magnon BEC phase, to be given by

\begin{equation}
\mu_{\mathrm{c=2/c=1}} =  \left(1+\frac{\alpha_\mathrm{eff}}{c \cdot \alpha_\mathrm{sp}}\right)\left[\hbar \gamma\mu_0 \left(H + \frac{M_\mathrm{s}}{2}\right)\right].
\label{eq:threshold}
\end{equation}
Here, $\mu_\mathrm{s} = \mu_{\mathrm{c=2}}$ corresponds to the critical spin chemical potential for the formation of a magnon BEC, while $\mu_\mathrm{s} = \mu_{\mathrm{c=1}}$ corresponds to the swasing instability, which is equivalent to the full damping compensation~\footnotemark[1]. Furthermore, $\alpha_\mathrm{eff}$ is an effective damping parameter~\footnotemark[1]
, $\alpha_\mathrm{sp}$ is the spin pumping induced damping enhancement of the FMI, $\hbar$ is the reduced Planck constant, $\gamma$ is the gyromagnetic ratio and $\mu_0$ is the vacuum permeability. The criteria further depend on the external magnetic field magnitude $H$ and the saturation magnetization $M_\mathrm{s}$. The spin chemical potential is related to the applied current by $\mu_\mathrm{s} = [{e \theta_\mathrm{SH} I_\mathrm{dc} }\tanh{\left(\eta\right)}]/[{w \sigma_\mathrm{e}\eta}] $~\cite{CornelissenTheory,ChenSMR,ZhangMMR2}, where $e$ is the elementary charge, $w$ denotes the width of the Pt strip, $\sigma_\mathrm{e}$ and $\theta_\mathrm{SH}$ are the electrical conductivity and the spin Hall angle of the Pt. Moreover, $\eta=t_\mathrm{HM}/(2 l_\mathrm{s})$ is the ratio of the thickness of the Pt strip $t_\mathrm{HM}$ and its spin diffusion length $l_\mathrm{s}$.   
For a comparison of our data to Eq.~\eqref{eq:threshold}, we plot the experimentally determined critical currents $I_\mathrm{on}$ and $I_\mathrm{crit}$ as a function of the applied magnetic field in Fig.~\ref{fig:threshold} (b). 
For  $I_\mathrm{on}$ ($I_\mathrm{crit}$), we observe a characteristic current around $\SI{0.45}{\milli\ampere}$ ($\SI{0.6}{\milli\ampere}$) for $\mu_0 H<\SI{50}{\milli\tesla}$ and both critical currents increase with the applied magnetic field strength for $\mu_0 H>\SI{50}{\milli\tesla}$.
We can solve the condition $\mu_{\mathrm{s}} = \mu_{\mathrm{c=2}}$ ($\mu_{\mathrm{s}} = \mu_{\mathrm{c=1}}$) for $I_\mathrm{dc}$ and identify the result with the aforementioned characteristic current $I_\mathrm{on}$ ($I_\mathrm{crit}$). Hence, we can quantitatively corroborate the field dependence of the critical currents observed in Fig.~\ref{fig:threshold} (b). Using the values $\sigma_\mathrm{e}=\SI{1.74e6}{1/\ohm\metre}$, $\theta_\mathrm{SH}=0.11$, $l_\mathrm{s}=\SI{1.5}{\nano\metre}$, $w=\SI{500}{\nano\metre}$ and $t_\mathrm{Pt}=\SI{3.5}{\nano\metre}$ to calculate $\mu_{\mathrm{s}}$, we find good quantitative agreement of model and experimental data for both $I_\mathrm{crit}$ (spheres and red line) and $I_\mathrm{on}$ (stars and blue line). 
The characteristic parameters $\alpha_\mathrm{eff}$ and $\alpha_\mathrm{sp}$ entering Eq.~\eqref{eq:threshold} are determined independently using ferromagnetic resonance experiments presented in the SI~\footnotemark[1]. The strong increase of the theoretically predicted threshold currents at small magnetic fields is not properly reflected by the experimental data. As discussed in Ref.~\footnotemark[1], however, this may be caused by an in-plane magnetocrystalline anisotropy field, e.g. due to the cubic anisotropy of our YIG film. 

For an intuitive understanding of the BEC and swasing scenario excited using spin Hall physics, we refer to Ref.~\cite{dcmagnonBEC}. Here, the threshold of the BEC is determined by the presence of a finite population of magnons in the ground state, corresponding to a phase transition of second order. In contrast, the swasing threshold is associated with the full compensation of the intrinsic damping and can be identified with a coherent magnetization precession. The difference between those  threshold values originates from the fact that magnons are an excitation with a finite lifetime and hence a non-conserved quantity. The observation of a smooth transition of $R_\mathrm{YIG}^\mathrm{s}$ in Fig.~\ref{fig:threshold} (a) thus might be indicative of this second order phase transition, where magnons are condensing continuously into a steady state BEC.

\textit{Summary} - We find ultra-low magnon resistance indicating an effectively vanishing magnon decay in a HM/FMI bilayer under the application of a large current density to the HM. The damping compensation is achieved by employing spin-orbit torque mediated spin current injection in a YIG/Pt heterostructure. 
We discuss our data by comparing it to the theoretically predicted threshold conditions for the transition into a DC charge current pumped magnon BEC~\cite{dcmagnonBEC}. 
This work lays the foundation for experiments ranging from zero resistance magnon transport to efficient non-linear spin current manipulation. 


\begin{acknowledgments}
	
	T.W. acknowledges Akashdeep Kamra for valuable discussions. This work is financially supported by the DFG via Germany's Excellence Strategy EXC-2111-390814868 as well as projects AL2110/2-1, WE5386/4-1 and HU1896/2-1. 
	
\end{acknowledgments}

\end{document}